\documentclass[prl,twocolumn,aps]{revtex4}

\begin{document}

\title{Reply to the Comment on \textquotedblleft Anyonic Braiding in Optical
Lattices\textquotedblright\ }
\author{Chuanwei Zhang$^{1}$}
\author{V. W. Scarola$^{1}$}
\author{Sumanta Rewari$^{1}$}
\author{S. Das Sarma$^{1}$}
\affiliation{$^{1}$Condensed Matter Theory Center, Department of Physics, University of
Maryland, College Park, MD 20742}

\begin{abstract}
This is the reply to the comment arXiv:0801.4620 by Vidal, Dusuel, and
Schmidt.
\end{abstract}

\maketitle

\bigskip In a Comment \cite{Vidal} on our recent work \cite{Zhang} on
anyonic braiding in optical lattices, Vidal \textit{et al.} claim that the
vanishing of certain spin correlators apparently invalidates our conclusions
in \cite{Zhang}. This claim \cite{Vidal} demonstrates a complete lack of
understanding of our work in particular and cold atom optical lattice
physics in general. We assert that all our results and conclusions in \cite%
{Zhang} remain valid in spite of the vanishing of certain spin correlation
functions \cite{Baskaran}. The main result derived in our work \cite{Zhang}
is an explicit experimental scheme involving external laser configurations
which is capable of carrying out anyonic braiding in optical lattices, and
nothing in the Comment by Vidal et al. \cite{Vidal} pertains to this central
result of our work.

The fact that the spin-spin correlation function $\left\langle \sigma
_{D^{\prime }}^{x}\sigma _{F}^{x}\right\rangle $ along a z-link is zero, as
found in ref. \cite{Baskaran}, only implies that our proposed technique for
detecting anyonic braiding needs to be applied to some other non-zero
spin-spin correlation function, as was already appreciated in ref. \cite%
{Baskaran}. For example, the non-zero function $\left\langle \sigma
_{D^{\prime }}^{x}\sigma _{V}^{x}\right\rangle $ along a x-link \cite%
{Baskaran} serves our purpose equally well, and can be used for detecting
anyonic statistics following the experimental scheme outlined in our paper.
We emphasize the fact, completely missed in \cite{Vidal}, that our work
provides a general experimentally feasible scheme for measuring arbitrary
spin correlators on optical lattices, which can be suitably adapted for
detecting anyonic statistics.

For the other comment concerning fermionic excitations \cite{Vidal}, we
merely point out that the fermionic excitations are protected by an energy
gap $2\left\vert J_{z}\right\vert $, which is much larger than the anyonic
excitation gap $J_{eff}$ in the relevant part of the phase diagram \cite%
{Kitaev}. When the spin operations are slower than $\sim \hbar /2\left\vert
J_{z}\right\vert $, the number of these excitations is exponentially small.

Finally, we mention that the task of an experimental demonstration of
anyonic braiding statistics in a real system, the goal of our work in ref.
\cite{Zhang}, is quite distinct from the understanding and the elucidation
of anyonic excitations in a well-defined theoretical model (e.g. the Kitaev
model \cite{Kitaev}), a distinction completely lost in the Comment by Vidal
\textit{et al.} \cite{Vidal}.

\end{document}